\documentclass[prb,twocolumn,showpacs,preprintnumbers,amsmath,amssymb,floatfix]{revtex4}

\usepackage{epsfig}
\usepackage{bm}

\begin{document}

\title{Spin-Peierls phases in pyrochlore antiferromagnets}

\author{Oleg Tchernyshyov}
\email{otcherny@princeton.edu}
\affiliation{Physics Department, Princeton University, 
Princeton, New Jersey 08544}
\author{R. Moessner}
\email{moessner@lpt.ens.fr}
\affiliation{Laboratoire de Physique Th\'eorique de l'Ecole Normale 
Sup\'erieure; CNRS-UMR 8541; 24, rue Lhomond; 75231 Paris Cedex 05; France}
\author{S. L. Sondhi}
\email{sondhi@princeton.edu}
\affiliation{Physics Department, Princeton University, 
Princeton, New Jersey 08544 }
\date{\today}

\begin{abstract}
In the highly frustrated pyrochlore magnet spins form a lattice
of corner-sharing
tetrahedra. We show that the tetrahedral ``molecule'' at the heart of
this structure undergoes a Jahn-Teller distortion when lattice motion
is coupled to the antiferromagnetism. We extend this analysis to the
full pyrochlore lattice by means of Landau theory and argue that it
should exhibit ``spin-Peierls'' phases with bond order but no spin
order. We find a range of N\'eel phases, with collinear, coplanar and
noncoplanar order.  While collinear N\'eel phases are easiest to
generate microscopically, we also exhibit an interaction that gives
rise to a coplanar state instead.
\end{abstract}

\maketitle


\section{Introduction}
\label{sec:intro}
The study of highly frustrated magnets began with Wannier and
Houtappel's realization that the triangular lattice Ising
antiferromagnet is paramagnetic at any nonzero temperature and
exhibits a macroscopic entropy even in at zero
temperature.\cite{Wannier,Houttapel} This canonical example
illustrates the defining characteristics of such systems - their
failure to order at temperatures of order the exchange constant,
empirically derivable from the high temperature Curie susceptibility,
and a large low-temperature entropy.\cite{SR,Moessner}

The advent of the cuprate superconductors led to seriously
renewed interest in these systems in the hope of finding
a quantum spin liquid - the zero temperature state of a quantum
magnet that fails to order. Subsequently their study has
blossomed, driven by an increasing list of materials that
exhibit highly frustrated antiferromagnetism, and is driven
as much in hopes of finding unusual ordering at low
temperatures. An appealing, if optimistic, analogy is to
the quantum Hall system, where the magnetic field frustrates
the kinetic energy and produces a macroscopic degeneracy,
which is then lifted by residual terms in the Hamiltonian
to produce a rich phase diagram with various orderings.

\begin{figure}
\epsfig{file=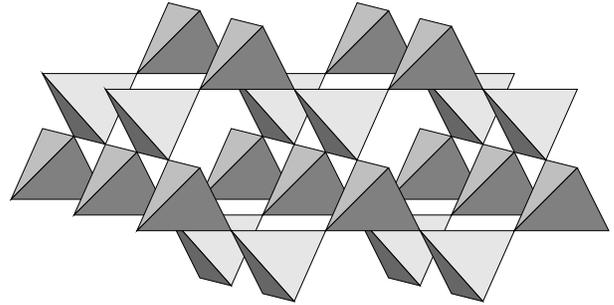,width=\columnwidth}
\caption{The ``pyrochlore'' lattice.  Magnetic ions are situated in
the corners of tetrahedra.}
\label{fig:lattice}
\end{figure}

The most promising system in this regard is the nearest-neighbor
Heisenberg system on the ``pyrochlore'' lattice, a network of
corner-sharing tetrahedara (Fig~\ref{fig:lattice}). The idealized
system has a vast ground state degeneracy in the classical limit of
infinite spin and there is a large and growing list of materials that
approximate this to varying degrees, including doped variants that
superconduct or display behavior reminiscent of the heavy fermions.

In this paper we limit ourselves to the insulating magnets.
Here it is known\cite{MC} that the classical system does {\em not}
exhibit order by disorder and remains a (co-operative)
paramagnet down to $T=0$. On general grounds one expects
that quantum fluctuations will select an ordering at sufficiently
large spin in a spin-wave treatment about the classical 
ground states.\cite{henley-unpub} At smaller values of spin,
the situation is unsettled with some form of singlet order
likely.\cite{HBB,CL}

Recently, building on work of Yamashita and Ueda,\cite{Ueda} we have
shown that in the presence of a coupling to the lattice,
a different mechanism of degeneracy lifting is likely to
operate.\cite{TMS} This involves
a version of the Jahn-Teller effect in which the lattice 
distorts to gain exchange energy (the ``spin-Teller'' effect)
and thereby relieves the frustration. 
As magnetoelastic couplings are ubiquitous, and lead to
a transition even at infinite spin, this mechanism will
dominate over the purely quantum selection effect, likely
starting at modest values of the spin.\cite{endnote1}
Also noteworthy in this problem is the likelihood of a finite 
temperature bond ordered phase preceeding the eventual establishment 
of Neel order. 

In this paper we present a detailed account of our anaysis of the
Jahn-Teller effect for Heisenberg magnets on the pyrochlore
lattice. Parts of this work have already been summarised in a short
paper.\cite{TMS} In Section \ref{sec:jt} we begin with the symmetry
analysis of the Jahn-Teller distortion of a single tetrahedron in the
classical limit and then extend it to ${\bf q}=0$ phonons for the
infinite lattice. Having identified the order parameter for lattice
distortions (bond ordering) in this fashion, in Section
\ref{sec:landau} we construct a Landau theory of the transition into
the bond ordered state which we contrast with the Landau theory of the
spin-Peierls transition in quasi-one dimensional systems. Finally we
turn to the establishment of N\'eel order which we discuss with the
insight gained from analyzing bond order (Section
\ref{sec:Neel}). Such ordering is most naturally collinear but
symmetry permits coplanarity and in Appendix \ref{sec:coplanar} we
describe an interaction that would bring it about. Appendix
\ref{sec:quantum} gives the quantum theory of the Jahn-Teller effect
in a single tetrahedron. The experimental situation with regard to
structure is briefly discussed in our concluding remarks in Section
\ref{sec:conclusion}. We will discuss the dynamical signatures of
various phases in a forthcoming publication.


\section{Jahn-Teller effect}
\label{sec:jt}

\subsection{Single tetrahedron}
\label{sec:jt-tetrahedron}

The structural unit of the pyrochlore antiferromagnet is a tetrahedral
``molecule'' with four spins in the corners.  Its high symmetry and the
degeneracy of the ground state are the two prerequisites for the
Jahn-Teller effect: the tetrahedron is distorted in its ground state.
The tendency of individual tetrahedra to deform induces a coherent
distortion of the entire crystal.  We will describe the Jahn-Teller
effect for a single tetrahedron in detail to understand which aspects
are relevant for the description of the spin-Peierls effect on the
entire lattice.

The energy of four spins on a regular tetrahedron is 
\begin{equation}
E_0 = J \sum_{i<j} {\bf S}_i \cdot {\bf S}_j 
= \frac{J}{2} ({\bf S}_1 + {\bf S}_2 + {\bf S}_3 + {\bf S}_4)^2
- 2JS(S+1). 
\label{eq:H-tetrahedron}
\end{equation}
In a ground state, the total spin is 0.  For quantum spins of length
$S$, there are $2S+1$ linearly independent ground states, which can be
constructed as follows.  The total spin of the pair ${\bf S}_1$ and
${\bf S}_2$ can be $0, 1, 2,\ldots,2S$, and likewise for the other
pair, ${\bf S}_3$ and ${\bf S}_4$.  An overall spin singlet can be
formed by combining two singlets, two triplets, and so on, giving a
total of $2S+1$ physically different singlet states.

The problem of quantum spins on an elastic tetrahedron can be solved
straightforwardly and is treated in detail in Appendix \ref{sec:quantum}. 
As the degeneracy and hence the Jahn-Teller distortion survive
at arbitrarily large values of spin, the outcome (with the
exception of the extreme quantum case $S=1/2$)  can be 
understood by looking at the simpler problem with classical spins.
In essence {\em this} spin-Peierls effect is {\em classical}, in
contrast with the usual cases where it goes away in that limit.

\begin{figure}
\begin{center}
\epsfig{file=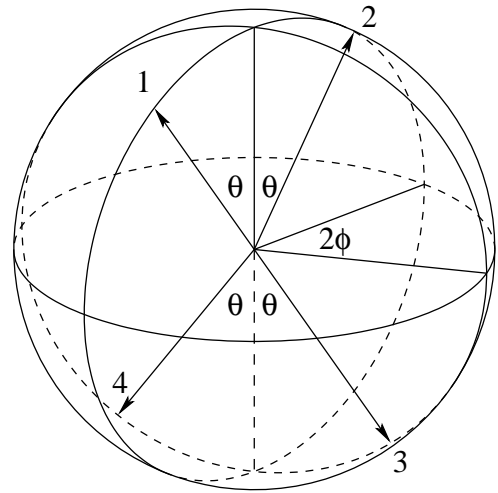,width=0.75\columnwidth}
\end{center}
\caption{Four spins of a tetrahedron in a ground state (zero total 
spin).}
\label{fig:sphere}
\end{figure}

For classical spins ($S \to \infty$), the degeneracy of the ground
state becomes continuous.  In addition to a trivial rigid rotation of
all four spins, there are two parameters that can be used for
characterization of a ground state (Fig.~\ref{fig:sphere}): the angle
$2\theta$ between spins 1 and 2 and the angle $2\phi$ between the
planes 12 and 34.  These two parameters determine the {\em bond variables}
in a ground state, 
\begin{eqnarray}
{\bf S}_1 \cdot {\bf S}_2 = {\bf S}_3 \cdot {\bf S}_4 
&=& S^2 \cos{2\theta},
\nonumber\\
{\bf S}_2 \cdot {\bf S}_3 = {\bf S}_1 \cdot {\bf S}_4 
&=& S^2 (\sin^2{\theta}\cos{2\phi} - \cos^2{\theta}),
\label{eq:opposites-equal}
\\
{\bf S}_3 \cdot {\bf S}_1 = {\bf S}_2 \cdot {\bf S}_4 
&=& -S^2 (\sin^2{\theta}\cos{2\phi} + \cos^2{\theta}).
\nonumber
\end{eqnarray}

At the heart of the effect lies the dependence of exchange interaction
on the relative positions of spins.  For example, if $J_{ij}$ depends
strictly on the distance between spins $i$ and $j$, the contribution
of this pair to exchange energy,
\[
E_{ij} = [J + (dJ/dr) \delta r_{ij} + \ldots]({\bf S}_i \cdot {\bf S}_j),
\]
generally has a term linear in the relative displacement $\delta
r_{ij}$.  Therefore, spins $i$ and $j$ exert on each other a force
$-(dJ/dr)({\bf S}_i \cdot {\bf S}_j)$, which is repulsive or
attractive depending on the angle between the spins.  In a generic
ground state (Fig. \ref{fig:sphere}), angles between spins are
unequal, so that disparate forces cause a deformation of the
tetrahedron.

More generally, exchange interaction may depend not only on the
distances between spins, but also on the angles between the bonds.  We
therefore write the magnetic and elastic energies of the spins in the
most general form:
\begin{equation}
E = E_0 + \sum_{a,i,j} (\partial J_{ij} / \partial x_a)
({\bf S}_i \cdot {\bf S}_j) x_a + \sum_{a,b} k_{ab} x_a x_b/2.   
\end{equation}
Here $E_0$ is the energy of a ground state, $x_1\ldots x_{12}$ are
Cartesian coordinates of the spins and $k_{ab}$ are the appropriate
elastic constants.  To reduce the number of independent coordinates
and forces, it is convenient to classify them in terms of irreducible
representations of the symmetry group $T_d$ of the tetrahedron.  Using
the appropriate linear combinations of coordinates $x_{\rho\alpha}$ and
forces $f_{\rho\alpha}$ (where $\rho$ labels irreducible representations, 
$\alpha$ enumerates its components) one obtains a simpler result:
\begin{equation}
E = E_0 + \sum_{\rho,\alpha} \left[ - J_\rho' 
f_{\rho\alpha} x_{\rho\alpha} + k_\rho x_{\rho\alpha}^2/2
\right].  
\end{equation}

\begin{figure}
\epsfig{file=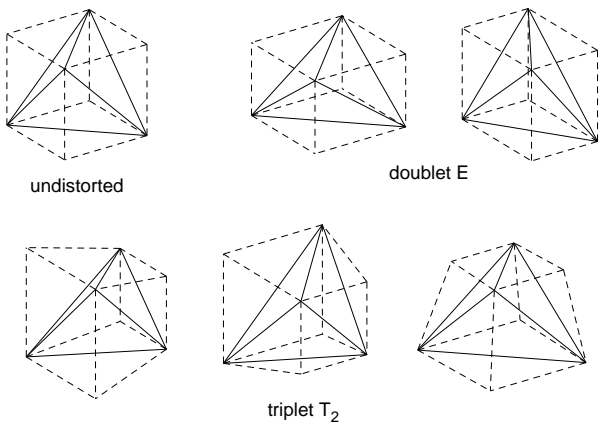,width=0.95\columnwidth}
\caption{Vibrational modes of a tetrahedral molecule.}
\label{fig:phonons}
\end{figure}

Six vibrational modes may affect the exchange energy: a singlet $A_1$,
a doublet $E$, and a triplet $T_2$.  The breathing mode $A_1$
uniformly rescales exchange interactions on all bonds and does not
discriminate between different ground states; therefore it can be left
out of consideration.  A component of the vector triplet $T_2$
stretches and contracts by the same amount two bonds opposite each
other (Fig.~\ref{fig:phonons}).  As can be inferred from
Eq. (\ref{eq:opposites-equal}), such bonds are equally satisfied (or
equally frustrated) in any ground state.  Therefore stretching one and
contracting the other to the same extent cancels the linear term in
magnetic energy, making the triplet mode ineffectual in relieving
frustration via the Jahn-Teller mechanism.  
Only one irreducible representation causes the
Jahn-Teller effect: the doublet $E$ whose components are tetragonal
and orthorombic distortions of the tetrahedron
(Fig.~\ref{fig:phonons}).  Since no other representations will be
dealt with, we will suppress the representation subscript $\rho = E$
in what follows.

\begin{figure}
\begin{center}
\epsfig{file=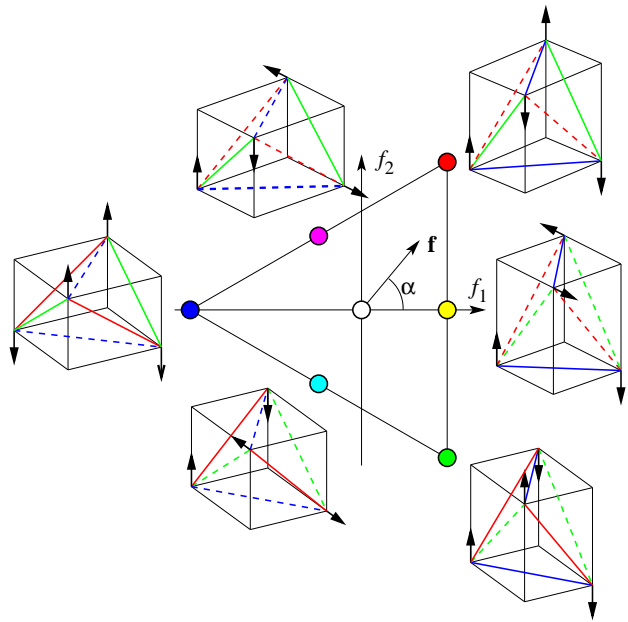,width=0.95\columnwidth}
\end{center}
\caption{The domain of the bond vector ${\bf f} = (f_1,f_2) =
(f\cos{\theta}, \, f\sin{\theta})$ is bounded by an equilateral
triangle in the $(f_1,f_2)$ plane.  Also shown are six extremal spin
configurations.  Strong (weak) bonds are denoted by solid (dashed)
lines.  The color of a state is determined by the color of
frustrated bonds.}
\label{fig:f}
\end{figure}

The six bond variables ${\bf S}_i \cdot {\bf S}_j$ contain the same
representations.  The singlet $A_1$ is the symmetric sum, which is
nothing but the energy of the undistorted ground state
(\ref{eq:H-tetrahedron}), i.e., a constant that does not favor any
particular ground state.  The triplet $T_2$ contains the differences
of forces on {\em opposite} bonds,
\begin{equation}
{\bf S}_1 \cdot {\bf S}_3 - {\bf S}_2 \cdot {\bf S}_4, \hskip 5mm
{\bf S}_1 \cdot {\bf S}_4 - {\bf S}_2 \cdot {\bf S}_3, \hskip 5mm
{\bf S}_1 \cdot {\bf S}_2 - {\bf S}_3 \cdot {\bf S}_4.
\label{eq:T2}
\end{equation}
As already mentioned, these differences vanish in a ground state.  The
remaining forces form a doublet $E$ showing the disparities between 
adjacent bonds: 
\begin{eqnarray}
f_1 &\!=\!& 
\big[({\bf S}_1 + {\bf S}_2)\!\cdot\!({\bf S}_3 + {\bf S}_4)
 - 2 {\bf S}_1 \cdot {\bf S}_2 - 2{\bf S}_3 \cdot {\bf S}_4 \big]/\sqrt{12},
\nonumber
\\
f_2 &\!=\!& 
({\bf S}_1 - {\bf S}_2) \cdot ({\bf S}_3 - {\bf S}_4)/2. 
\label{eq:f}
\end{eqnarray}
The component $f_1$ shows by how much the blue bonds are stronger than the
rest (Fig. \ref{fig:f}); $f_2$ measures the difference between the red
and green bonds.  The domain of possible values of the vector ${\bf f}
= (f_1, f_2) = (f\cos{\alpha},f\sin{\alpha})$ is an equilateral
triangle.  Its perimeter is made of coplanar ground states; the three
corners correspond to the three distinct collinear ground states; they
are marked by the color of frustrated bonds.  The two components of
${\bf f}$, like the two angles in Fig.~\ref{fig:sphere}, can be used
to parametrize degenerate classical ground states of a tetrahedron; in
fact, $f_1 = 2S^2(1-3\cos^2{\theta})/\sqrt{3}$, $f_2 =
2S^2\sin^2{\theta}\cos{2\phi}$.

After these simplifications, the energy of the system has the form
\begin{equation}
E = E_0 - J' {\bf f \!\cdot\! x} + k x^2/2,
\label{eq:e-x-f}
\end{equation}
where ${\bf x} = (x_1, x_2)$ are amplitudes of the tetragonal and
orthorombic distortions.  $J'$ and $k$ are the magnetic and elastic
constants appropriate for the $E$ representation.  The energy is
minimized when $k{\bf x} = J' {\bf f}$, so that
\begin{equation}
E_{\rm min} = E_0 - {J'}^2 f^2/2k.
\label{eq:f-squared}
\end{equation}
One can view the $-f^2$ term as a quartic spin interaction
\begin{equation}
- \frac{{J'}^2 f^2}{2k} 
= -\frac{{J'}^2}{3k} \sum_{i>j} ({\bf S}_i \cdot {\bf S}_j)^2 
+ \mbox{ const}
\end{equation}
induced by ``integrating out'' the phonons.\cite{Kittel} It evidently
prefers collinear ground states with four maximally satisfied bonds
and two maximally frustrated bonds.  The resulting distortion of the
tetrahedron is tetragonal.  It flattens or elongates along one of its
$C_2$ axes, depending on the sign of the derivative $J'$.

Modulo global rotations of the spins, there are three degenerate
ground states, which we label with the colors 
red, blue and green according to which pair of opposite bonds is frustrated (Fig.~\ref{fig:f}).  Their opposites are cyan,
magenta, and yellow states with four frustrated bonds.  We have found
that ground states can have these secondary colors in a model with
more general spin interactions, e.g., 4-spin cyclic exchanges.  See
Appendix \ref{sec:coplanar} for details.

A consideration of quantum spins in the Appendix \ref{sec:quantum} yields
essentially the same result: the energy is minimized when two opposite
bonds (e.g., 12 and 34) have the highest spins $2S$ each.  Such states
are the quantum analogue of parallel spins.  In contrast to the
spin-Peierls effect on a Heisenberg chain, this one is a classical
affair: instead of forming spin singlet on stronger bonds, Heisenberg
spins of a tetrahedron form the highest spin on two weak bonds.

\begin{figure}
\epsfig{file=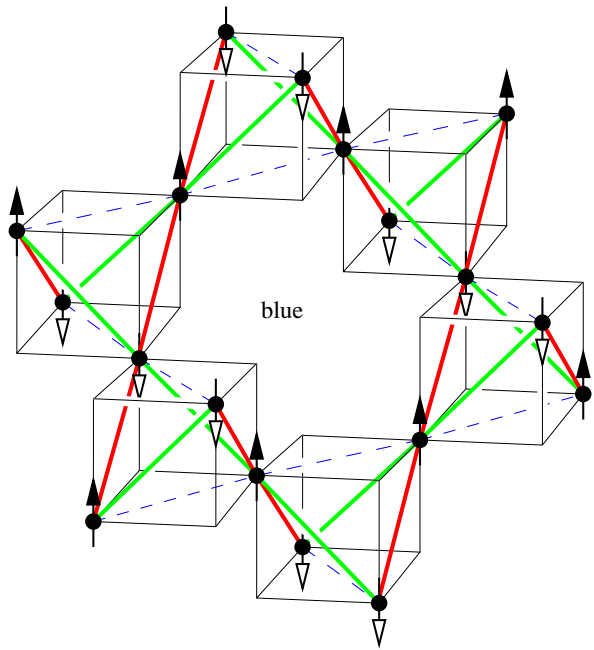,width=0.9\columnwidth}
\caption{The N\'eel state obtained in the magnetoelastic model with
${\bf q} = 0$ phonons.  The $E_g$ phonon mode dominates.  The
distortion weakens blue bonds on all tetrahedra (color scheme of
Fig.~\ref{fig:f}).}
\label{fig:Neel-g}
\end{figure}

\subsection{Pyrochlore lattice: ${\bf q} = 0$ phonons}
\label{sec:jt-lattice}

An attempt to extend this calculation to an infinite network of
tetrahedra runs into a substantial problem: all possible phonon modes,
the number of which is proportional to the number of tetrahedra,
couple to bond variables, and any one of them may thus lead to a
magnetoelastic distortion. In order to describe the basic physics of
the Jahn-Teller effect on the full lattice, in this section we
restrict ourselves to the (over)simplified
version with only a few participating phonons.  As a
result of this crude approximation, ``integrating out'' the phonons
produces an infinite-range interaction between vectors ${\bf f}$ of
different tetrahedra.  In a realistic model including phonons of all
wavelengths, such forces will have a finite radius.  However, the
structure of the ground state is often insensitive to such details.

\begin{figure}
\epsfig{file=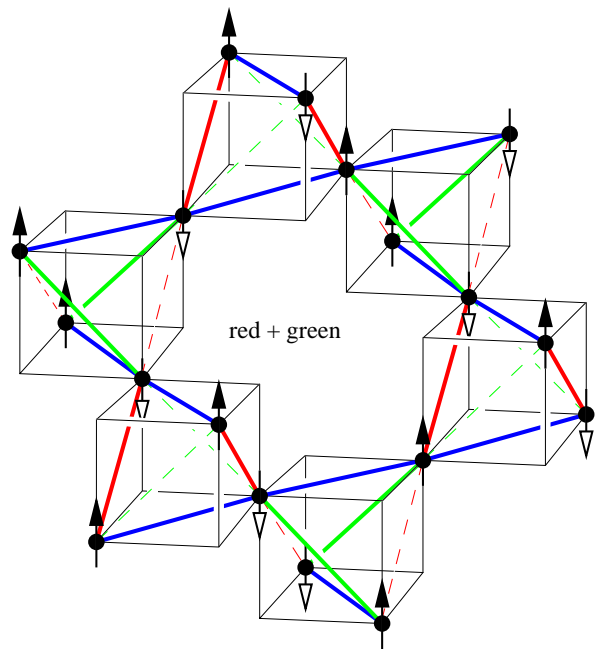,width=0.9\columnwidth}
\caption{Same as Fig.~\ref{fig:Neel-g} but with the dominant $E_u$
phonon.  Red and green bonds are weakened alternatively.}
\label{fig:Neel-u}
\end{figure}

We specialize to the case of phonons with the wavevector ${\bf q} =
0$.  In effect, this restricts all tetrahedra of the same orientation
to have the same distortion (Figs.~\ref{fig:Neel-g} and
\ref{fig:Neel-u}).  The existence of two types of tetrahedra (labeled
in what follows A and B), which differ by orientation, is the only new
degree of freedom.  The symmetry group of the lattice (with equivalent
tetrahedra identified) is extended from $T_d$ to $I \times T_d \equiv
O_h$ by the operation of inversion on any site, which exchanges
tetrahedra A and B.  Irreducible representations are those of $T_d$
labeled by an additional quantum number, parity under the inversion.
The relevant phonons are $E_g$ and $E_u$, which are, respectively,
uniform or staggered distortions of the lattice.  For example, the
first component of $E_g$ stretches all tetrahedra along the $z$
direction (resulting in a macroscopic distortion of the crystal),
whereas the first component of $E_u$ stretches tetrahedra A and
squeezes tetrahedra $B$ along the same axis (leaving the crystal
dimensions unaltered to leading order).  The resulting distortions of
tetrahedra A and B can be written
\[
{\bf x}_A  = ({\bf x}_g + {\bf x}_u)/\sqrt{2}, 
\hskip 5mm
{\bf x}_B  = ({\bf x}_g - {\bf x}_u)/\sqrt{2}.
\]
The sum of elastic and magnetic energies,
\[
E = - J'({\bf f}_A \cdot {\bf x}_A + {\bf f}_B \cdot {\bf x}_B)
+ \frac{k_g|{\bf x}_g|^2}{2} 
+ \frac{k_u|{\bf x}_u|^2}{2},
\]
is readily minimized with respect to the phonon variables:
\begin{eqnarray}
E_{\rm min} = - \frac{J'|{\bf f}_A + {\bf f}_B|^2}{4k_g} 
- \frac{J'|{\bf f}_A - {\bf f}_B|^2}{4k_u}.  
\label{eq:E-g-u}
\end{eqnarray}
The minimized energy thus consists of two terms.  The first term is diagonal 
in ${\bf f}_A$ and ${\bf f}_B$:
\[
-J' \left(k_g^{-1} + k_u^{-1}\right) (f_A^2 + f_B^2)/4.
\]
It puts tetrahedra of both types into one of the three collinear states
(red, green, or blue), thus defining a 3-state Potts model.  The
cross term,
\[
-J' \left(k_g^{-1} - k_u^{-1}\right) ({\bf f}_A \cdot {\bf f}_B)/2,
\]
introduces a coupling between the Potts states on the two sublattices.
A softer even phonon ($k_g < k_u$) yields the ground state of a 
ferromagnetic Potts model: all tetrahedra have the same primary color.
A softer odd phonon ($k_g > k_u$) produces a ground state with two
different primary colors on the two sublattices.  

Translated back into spin language, the two ground states are shown in
Figs. \ref{fig:Neel-g} and \ref{fig:Neel-u}.  The latter, in fact,
describes the N\'eel state observed in YMn$_2$ and MgV$_2$O$_4$,
compounds with spontaneous structural distortions.


\section{Landau theory}
\label{sec:landau}

Our simple model of classical spins on an elastic lattice of tetrahedra
appears to be reasonably successful in explaining ground-state properties
of some frustrated magnets.  Can we also gain some understanding of phase
transitions in these materials?  

To start with, we need to identify the relevant phases.  At high
temperatures, we have a symmetric paramagnetic state with no spin
or bond 
order and no lattice distortions.  The ground state ($T=0$) is a
N\'eel phase with a distorted lattice.  The two phases are
distinguished, for example, by spin averages $\langle {\bf S}_i
\rangle$ and by the disparities in bond lengths.  
In general, there may (and in certain cases will) exist an
intermediate spin-Peierls phase.  It is distinct from the N\'eel phase
by the absence of spin order ($\langle {\bf S}_i \rangle = 0$).  It is
also different from the paramagnetic phase by the presence of lattice
distortions and unequal spin correlations $\langle {\bf S}_{i}
\!\cdot\! {\bf S}_{j} \rangle$ between various nearest neighbors, with
a concomitant lowered symmetry.  In
such cases, we expect two phase transitions: first, a high-temperature
spin-Peierls transition, which partially relieves frustration of
spins, then, at a lower temperature, a transition into a N\'eel state.
Such a scenario is permitted by symmetry and the frustration makes
it easier to generate fluctuations that destablize the N\'eel state
without destroying the bond order.

In this section, we will discuss the spin-Peierls transition using
Landau theory.  By analogy with spin chains, we will identify the
relevant order parameter and discuss possible phase transitions in the
framework of the Landau theory.

\begin{figure}
\epsfig{file=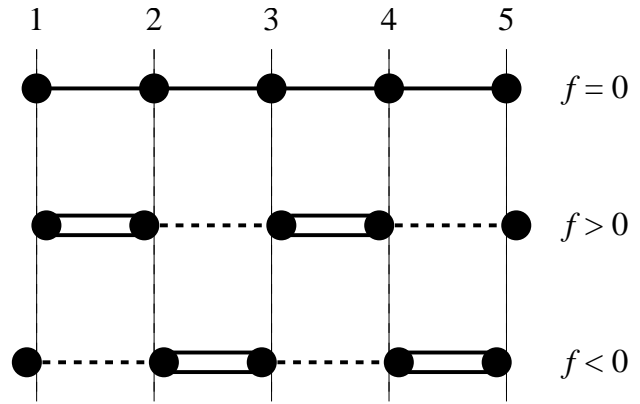,width=0.95\columnwidth}
\caption{Spontaneous dimerization of a spin chain.}
\label{fig:chain}
\end{figure}

\subsection{Dimerized chain}
\label{sec:landau-chain}

To introduce method and notation, we start with a familiar example,
namely the Landau theory of the spin-Peierls
transition in antiferromagnetic chains coupled to three-dimensional
phonons.  The paramagnetic and
dimerized phases can be distinguished using the order parameter
\begin{equation}
f = \langle {\bf S}_{2n+1} \!\cdot\! {\bf S}_{2n} 
 - {\bf S}_{2n} \!\cdot\! {\bf S}_{2n-1} \rangle. 
\label{eq:f-chain}
\end{equation}
It vanishes in the paramagnetic phase, since all bonds are equivalent.
Spontaneous dimerization increases the probability of finding a
singlet on half of the bonds, which leads to a nonzero value of $f$.
Expansion of the free energy (per spin) in powers of the order
parameter contains even powers of $f$ only:
\begin{equation}
F(f,T) = F(0,T) + a(T) f^2 + c(T) f^4 + \ldots
\label{eq:f-chain-series}
\end{equation}
Assuming that $a$ becomes negative below $T=T_c$, so that $a(T)
\approx \alpha(T-T_c)$, while $c>0$ and roughly constant, one obtains
the standard scenario of a second-order phase transition: the minimum
of the free energy shifts continuously from $f=0$ above $T_c$ to $f =
\pm \sqrt{\alpha|T-T_c|/2c}$ below $T_c$.

The continuity of the transition depends crucially on the absence of 
a cubic term in the expansion (\ref{eq:f-chain-series}).  With chains,
its absence is guaranteed by symmetry: states with $f$ differing only
by a sign are physically equivalent (Fig.~\ref{fig:chain}), hence only
even powers of $f$ are allowed.  

Formally, the fate of an $f^3$ term is decided by its symmetry
properties.  The symmetry group of an undistorted chain includes
inversion on a site, which takes $f \mapsto -f$.  Likewise, $f^3
\mapsto -f^3$.  Since, however, free energy must be invariant under
all symmetry transformations, an $f^3$ term is forbidden.  

In contrast, we will find that a cubic term is allowed in certain
cases for the spin-Peierls order parameter on the pyrochlore lattice.
In such cases, the spin-Peierls transition is expected to be
discontinuous.

\subsection{Pyrochlore lattice: order parameter and broken symmetries}

What order parameters would characterize a spin-Peierls phase in a
network of tetrahedra?  The smallest unit of the lattice, a
tetrahedron, contains 6 bond variables, so that there are 6 averages
$\langle{\bf S}_i \cdot {\bf S}_j\rangle$ and 5 differences between
them, all of which could serve as order parameters.  From the symmetry
viewpoint, they can be divided into irreducible representations of the
tetrahedron group.  One of them is the doublet ${\bf f} = (f_1, f_2)$,
where
\begin{eqnarray}
f_1 & = & 
\frac{\langle({\bf S}_1 + {\bf S}_2)\!\cdot\!({\bf S}_3 + {\bf S}_4)
 - 2 {\bf S}_1 \cdot {\bf S}_2 - 2{\bf S}_3 \cdot {\bf S}_4 \rangle}
{\sqrt{12}},
\nonumber
\\
f_2 & = & 
\frac{\langle({\bf S}_1 - {\bf S}_2) \cdot ({\bf S}_3 - {\bf S}_4)\rangle}{2}.
\label{eq:f-average}
\end{eqnarray}
The other is a triplet - see Eq.~(\ref{eq:T2}).  In the paramagnetic
phase, both the doublet and triplet order parameters must vanish (all
nearest-neighbor bonds have the same strength).  In a spin-Peierls
phase, either the doublet, or the triplet (or, potentially, both) will
have nonzero expectation values.  Energy considerations of Section
\ref{sec:jt} suggest that the driving force of this transition is the
doublet.

The two-component order parameter ${\bf f}$ can be the same for all
tetrahedra, in which case only the rotational symmetry of the lattice
will be broken.  Symmetry with respect to inversion on a site can also
be violated if the order parameter ${\bf f}$ is not the same on
tetrahedra of different orientations.  Lastly, translational symmetry
of the lattice can also be broken if ${\bf f}$ varies among equivalent
tetrahedra forming a commensurate wave.  

\subsection{Pyrochlore lattice: ${\bf q} = 0$ phonons}
\label{sec:landau-lattice}

We restrict the analysis to situations when the translational symmetry
of the lattice remains intact, as we did previously in Section
\ref{sec:jt-lattice}.  In this case, any two tetrahedra of the same
orientation distort in the same way reducing the space group of the
pyrochlore lattice to the octahedral point group $O_h \equiv I \times
T_d$ (inversion $I$ exchanges tetrahedra of different orientations).
Despite this rather drastic simplification, we will see that we can
account for the experimentally observed behaviour of a number of
compounds, at least qualitatively.  Phonons at other points in the
Brillouin zone may drive a spin-Peierls transition as well, 
leading to ordered states with larger and often more complex unit cells.

The order parameter has (potentially unequal) values ${\bf f}_A$ and
${\bf f}_B$ on tetrahedra of inequivalent orientations.  Their
symmetric and antisymmetric combinations ${\bf g} = {\bf f}_A + {\bf
f}_B$ and ${\bf u} = {\bf f}_A - {\bf f}_B$ are irreducible doublets
of the group $O_h$.  In the paramagnetic phase, ${\bf g} = {\bf u} =
0$.  In various spin-Peierls phases, one or both of these order
parameters are nonzero.

For classical spins, the domain of possible values of the order
parameters ${\bf f} = (f_1, f_2)$ is the familiar triangle shown in
Fig.~\ref{fig:f}.  In view of the three-fold symmetry (more precisely,
permutation group $S_3$), the two-dimensional vector ${\bf f}$ can be
interpreted as a color.  The extremal points represent the red, blue,
and green states with collinear spins.  In the paramagnetic (white)
state ${\bf f} = 0$, the color symmetry is manifest: the three primary
colors are represented equally.  In any spin-Peierls phase, the global
color symmetry $S_3$ is spontaneously broken.

\subsubsection{Landau free energy}

The $O_h$ symmetry of the high-temperature phase allows the following
terms in the Landau free energy:
\begin{subequations}
\begin{eqnarray}
F({\bf g,u}) &=& a_g g^2 + b_g g^3\cos{3\theta_g} + c_g g^4 + \ldots
\label{GLa}\\
&+& a_u u^2 + c_u u^4 + d_u u^6 \cos{6\theta_u} + \ldots 
\label{GLb}
\\
&+& b_u u^2 g \cos{(2\theta_u+\theta_g)} + \ldots
\label{GLc}
\end{eqnarray}
\end{subequations}
Here ${\mathbf{g}}=(g \cos\theta_g,g \sin\theta_g)$ with an analogous
definition of $u$ and $\theta_u$. The first (second) line contains the
leading terms for the even (odd) distortion; the third line represents
the lowest-order coupling between ${\bf g}$ and ${\bf u}$. The
constants $a$ through $e$\ in this expression cannot be determined by
symmetry considerations alone; when convenient, one can try to
determine their likely sign by taking recourse to microscopic model
Hamiltonians for the spin-lattice system.

Omitted higher-order terms are assumed to be positive for stability.
Landau theory is of course strictly to be applied only for small
values of the order parameters. However, the shape of the range of the
vector ${\bf f}$\ (Fig~\ref{fig:f}) encodes some information on where
the order parameter, once close to saturation, may point.

This form of the free energy permits a number of distinct ordered
states.  Generally, the symmetry of the lattice is reduced from cubic
to tetragonal.  In addition, the presence of an odd distortion (${\bf
u} \neq 0$, or ${\bf f}_A \neq {\bf f}_B$) also breaks the symmetry of
inversion through a site, exchanging tetrahedra $A$ and $B$.  Note
that, whenever a staggered distortion ${\bf u}$ is present, the
coupling term (\ref{GLc}) generates a subdominant uniform distortion
${\bf g}$ of the crystal.

The phase transitions can be first or second order, depending on the
mode driving the transition: the free energy of the even mode ${\bf
g}$ may have a cubic term (\ref{GLa}), which generally leads to a
discontinuous jump.  The odd mode ${\bf u}$ does not have its own
cubic term (\ref{GLb}), but is instead coupled nonlinearly to ${\bf
g}$ (\ref{GLc}).  This difference has the following physical origin.
When an even distortion is present, all tetrahedra have, say, 4 strong
and 2 weak bonds (a state of primary color).  Changing the sign of the
order parameter ${\bf g}$ would give a state with 2 strong and 4 weak
bonds on every tetrahedron (secondary color), which need not have the
same free energy, $F({\bf g},0) \neq F(-{\bf g},0)$.  Hence a $g^3$
term is allowed.  On the other hand, in a state with a pure odd
distortion, tetrahedra A and B have opposite colors (e.g., A is blue
and B is yellow) and switching the sign of ${\bf u}$ merely exchanges
them (A is yellow and B is blue), so that $F(0,{\bf u}) = F(0,-{\bf
u})$.  Hence, there is no $u^3$ term.

\subsubsection{Spin-Peierls phases}
\label{section:landau-lattice-phases}

\begin{figure}
\epsfig{file=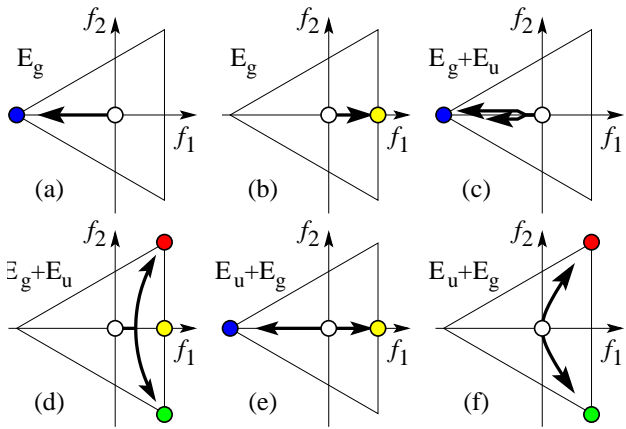,width=0.95\columnwidth}
\caption{Development of the order parameters ${\bf f}_A$ and ${\bf
f}_B$ in the six scenarios of the spin-Peierls phase transition.
${\bf f} = 0$ in the paramagnetic (white) state.  (a--b) A
first-order transition is driven by the $E_g$ phonon and is
discontinuous.  ${\bf f}_A = {\bf f}_B \neq 0$.  (c--d) As a result of
that transition, the $E_u$ phonon softens and triggers a subsequent
second-order transition into a phase with ${\bf f}_A \neq {\bf f}_B$.
(e--f) A second-order transition is driven by the $E_u$ phonon.  The
$E_g$ order parameter is also induced, so that ${\bf f}_A \neq -{\bf
f}_B$.}
\label{fig:GL}
\end{figure}

In the high-temperature paramagnetic phases, $a_g>0$ and $a_u>0$, the
minimum of the free energy lies at ${\bf g} = {\bf u} = {\bf f}_A =
{\bf f}_B = 0$.  At low enough temperatures, one or both of these
coefficients may become negative --- see Eq.~(\ref{eq:E-g-u}).  The
nature of the resulting phase transition depends sensitively on the
order in which $a_g$ and $a_u$ turn negative, as well as on the signs
of the Taylor coefficients $b_u$, $b_g$, $d_u$.  Our results for the
Heisenberg model (Section \ref{sec:jt}) are compatible with the choice
$b_u>0$, which we will assume in what follows.  
Below we describe six scenarios depicted in Fig.~\ref{fig:GL}.

(a) In the simplest case, the even mode ${\bf g}$ becomes unstable,
while the odd mode remains suppressed at all temperatures.  The
transition is discontinuous because of the cubic term in the free
energy (\ref{GLa}).  For $b_g>0$, minima of the free energy are at
$\theta_g = \pi, \pm\pi/3$.  As ${\bf u} = 0$, distortions are the
same on all tetrahedra, ${\bf f}_A = {\bf f}_B = {\bf g}/2$.  Thus
each tetrahedron shows the same tetragonal distortion with 4 strong and
2 weak bonds (one of the three primary-color states).

(b) Same as (a) but $b_g<0$.  The minima are at $\theta_g = 0, \pm
2\pi/3$, secondary-color states with a tetragonal distortion, 2 strong
and 4 weak bonds on all tetrahedra.  

(c) As the even order parameter ${\bf g}$ grows, it modifies the
quadratic term of the odd mode through the nonlinear coupling
(\ref{GLc}).  Once $a_u - b_u g$ vanishes, scenarios (a) and (b) are
modified: a second, continuous transition occurs into a state where
both ${\bf g} \neq 0$ and ${\bf u} \neq 0$, so that ${\bf f}_A \neq
{\bf f}_B$.  The directions of ${\bf g}$ and ${\bf u}$ are correlated:
$2\theta_u + \theta_g = \pi$.  For $b_g>0$ (a), ${\bf u}$ is parallel
to ${\bf g}$; therefore, vectors ${\bf f}_A$ and ${\bf f}_B$ still
point towards one of the corners of the triangle, but their length
differ.  Distortions of tetrahedra A and B are remain tetragonal, but
are unequal in strength.  The symmetry of inversion is broken.  

(d) When the odd mode softens for $b_g<0$ (b), ${\bf u}$ is
perpendicular to ${\bf g}$.  The uniformly distorted state (e.g.,
yellow) becomes nonuniform: tetrahedra A acquire a red component,
tetrahedra B acquire a green one.  Distortions of individual
tetrahedra are no longer purely tetragonal: there is an orthorombic
component.  Because the latter has a staggered nature, the lattice as
a whole retains tetragonal symmetry.  The symmetry of inversion is
broken.

Caveat.  Because the high-temperature transitions in cases (c) and (d)
are discontinuous, the intermediate phase (single primary or secondary
color) may be skipped completely.  In that case, instead of a
succession of two transitions, there will be a single, discontinuous
transition directly into the final state with two different colors,
${\bf f}_A \neq {\bf f}_B$.

(e) The transition can also be driven by the odd phonon, in which case
it is expected to be continuous.  The initial direction of the vector
${\bf u}$ is determined by the sign of the sixth-order anisotropy
$d_u$.  For $d_u<0$, $\theta_u = n\pi/3$, where $n$ is an integer;
vectors ${\bf f}_A$ and ${\bf f}_B$ point in opposite directions, say,
${\bf f}_A$ has a primary color (e.g., blue), while ${\bf f}_B$ has a
secondary one (in this case, yellow).  The nonlinear coupling term
(\ref{GLc}) generates a subdominant order parameter ${\bf g} = {\cal
O}(u^2)$ parallel to ${\bf u}$.  This parasitic order parameter
increases the length of the primary-color component (${\bf f}_A$
becomes a deeper blue) and reduces that of the secondary-color
component (${\bf f}_B$ is a pale yellow).  The distortions are
tetragonal but of oposite directions on tetrahedra A and B.  

(f) Odd phonon with $d_u>0$.  The free energy (\ref{GLb}) has a
minimum for $\theta_u = (2n+1)\pi/6$.  Initially, distortions of
tetrahedra A and B are orthorombic, e.g. along $f_2$ (a red or green,
respectively, with a touch of blue).  The parasitic component ${\bf g}
= {\cal O}(u^2)$ and perpendicular to ${\bf u}$, bends ${\bf f}_A$ and
${\bf f}_B$ towards the primary red and green directions and makes the
individual distortions mostly tetragonal (along orthogonal axes in
real space).  Note that the final state is the same as in (d).

\subsection{Relation to 3-state Potts models}
\label{sec:landau-potts}

As we have already mentioned, the symmetry of the bond variables ${\bf
f}$ (permutation group $S_3$) invokes a similarity to the Potts model
with $q = 3$ states\cite{Potts} with energy 
\begin{equation}
E = J \sum_{\langle ij \rangle} \delta_{s_i s_j},
\end{equation}
$s_i = 1, 2, 3$ being Potts states.  Indeed, a similar two-component
order parameter has been introduced for the $q=3$ Potts model by
Ono.\cite{Ono} The pure Potts states correspond to collinear spin
configurations.  In the current context, the order parameter lives on
tetrahedra, which form a three-dimensional diamond lattice.  It is 
entirely plausible that the spin-Peierls transitions described in this
paper should be analogous to phase transitions in Potts models with 
short-range interactions.  

To this end, we can identify the simplest scenario
[Fig.~\ref{fig:GL}(a)] with the ferromagnetic Potts model.  The latter
is known to have a first-order transition in three
dimensions,\cite{Potts} which is consistent with our mean-field
result.  

Transitions shown in Figs. ~\ref{fig:GL}(e) and (f) have their analogs
in the antiferromagnetic 3-state Potts model.  Results (mostly
numerical) for lattices in $d=3$ dimensions have been obtained fairly
recently.\cite{Banavar,Rosengren,Rahman} Banavar {\em et
al.}\cite{Banavar} have studied the model on the simple cubic lattice
and found at low temperatures an ordered state with a broken
sublattice symmetry (BSS).  As the name suggests, the two sublattices
are inequivalent: spins on one sublattice are primarily in one Potts
state (say, blue), while the other sublattice is dominated by the
remaining spin states (red + green) in equal proportions (yellow).
More recently, Rosengren and Lapinskas predicted the existence of
another phase, with permutationally symmetric sublattices (PSS), where
sublattices A and B are dominated by two different Potts
states,\cite{Rosengren} e.g., red and green, respectively.  Their
Monte Carlo simulations suggest that the 3-state Potts antiferromagnet
on the diamond lattice orders into the PSS phase.\cite{Lapinskas} In
both cases, the phase transition appears to be continuous, with
critical properties of the XY model in $d=3$.  


\section{N\'eel phases}
\label{sec:Neel}

The spin-Peierls transition, whether in chains or in three-dimensional
magnets, is driven by the desire of spins to reduce frustration.  In
the bond-ordered phase, exchange strength varies from bond to bond
because of the distortion.  Thus frustration is relieved and the
classical ground state becomes unique, modulo global spin rotations.
In three dimensions, we can expect a spin-ordered state at zero
temperature.  As argued before, the transition into a N\'eel state
need not coincide with the spin-Peierls transition.  Therefore,
generally there will be three separate phases: paramagnetic,
spin-Peierls, and N\'eel.  (In those cases when the spin-Peierls
transition is discontinuous, the system may go directly into the
N\'eel phase bypassing the spin-Peierls stage.)

\subsection{N\'eel orders}
\label{sec:Neel-orders}

Particulars of the N\'eel order on a distorted lattice obviously
depends on the details of the distortion, which strengthens some bonds
and weakens others.  Because precise knowledge of spin interactions is
rarely available (even for the undistorted state!), one can try an 
alternative route,  namely to include spin averages $\langle {\bf S}_i \rangle$
in the Landau theory developed above for a spin-Peierls phase.  

To keep technical details to a minimum, we will restrict the
discussion to N\'eel states that do not break translational symmetry
of the crystal, i.e., spin averages $\langle {\bf S}_i \rangle$ will
be assumed to be identical for all tetrahedra of the same orientation.
Put another way, $\langle {\bf S}_i \rangle$ is the average spin on
the $i$-th sublattice, $i=1\ldots4$.  Evidently, this parametrization
adequately describes only a fraction of possible antiferormagnetic
orders. For example, one of the N\'eel ground states obtained in our simple
magnetoelastic model (Fig.~\ref{fig:Neel-u}) is already beyond its
scope. More generally, even $q=0$\ bond ordered states with
different strengths
on the pair of bonds of
inequivalent tetrahedra related by inversion cannot be translated into
a $q=0$\ spin state.

\subsubsection{Undistorted lattice}

Let us construct the Landau free energy for spins on an undistorted
lattice.  Using the order parameters ${\bf s}_i = \langle {\bf S}_i
\rangle$ one obtains the following expansion for the free energy:
\begin{equation}
F(\{{\bf s}_i\}) 
= \frac{a_0}{4} \sum_{i=1}^4 s_i^2 
+ a_1 \left( \sum_{i=1}^4 {\bf s}_i \right)^2
+ \frac{b_0}{4} \sum_{i=1}^4 s_i^4 + \ldots
\label{eq:F-Neel}
\end{equation}
In an antiferromagnet, $a_1>0$; for stability, we take $b_0>0$.  When
$a_0$ becomes negative, the minimum of the free energy shifts away
from ${\bf s}_i = 0$ and the system will enter a N\'eel state.  The
free energy is minimized by {\em any} configuration of spin averages
satisfying $\sum_{i=1}^4 {\bf s}_i = 0$; the length of the averages is
given by $s_i^2 = -a_0/2b_0$.  The N\'eel pattern is thus not unique,
as expected for a frustrated magnet.

In addition to the quartic term shown in Eq.~(\ref{eq:f-Neel}), the free 
energy expansion may contain one more quartic invariant, 
\begin{equation}
b_1 \sum_{i>j}({\bf s}_i \!\cdot\! {\bf s}_j)^2.
\label{eq:biquad}
\end{equation}
This term, in fact, will break the degeneracy of the N\'eel states.
In the case $b_1<0$, the N\'eel phase has collinear spins (any one of
the three collinear states in Fig.~\ref{fig:f}.  Note that these
states also break the bond symmetry.)  When $b_1>0$, the spin averages point
at equal angles of $\arccos{(-1/3)} \approx 109^\circ$ to one
another.  

\begin{figure}
\epsfig{file=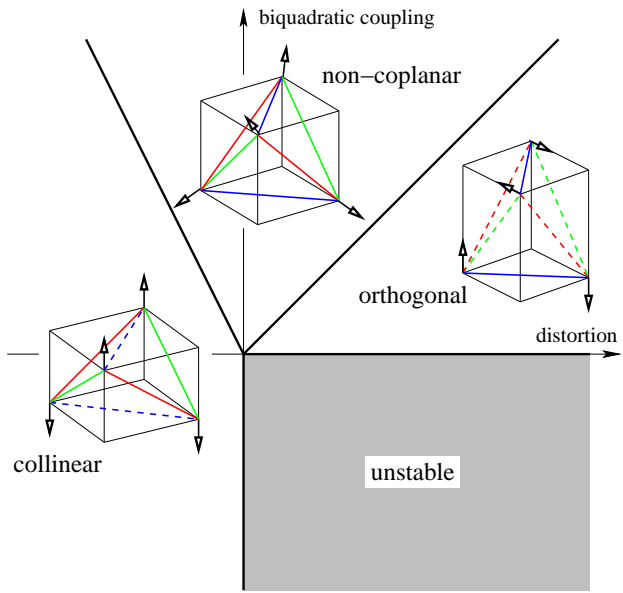,width=0.95\columnwidth}
\caption{N\'eel order in the presence of a tetragonal distortion, as
given by Landau theory.  The abscissa is the distortion amplitude 
$(a_2/a_0) f_1$; the ordinate is the biquadratic coupling $b_1/b_0$. 
The four N\'eel phases are described in the text.}
\label{fig:Neel-GL}
\end{figure}

\subsubsection{Distorted lattice}

A lattice distortion, however small, breaks the cubic symmetry, so that
additional, less symmetric terms will appear in the free energy.  For
a distortion that, symmetry-wise, belongs to the irreducible doublet $E$, 
the lowest-order perturbation will be of the form 
\begin{equation}
a_2 ({\bf f}\!\cdot\!{\bm \phi}) = a_2 (f_1 \phi_1 + f_2 \phi_2),
\label{eq:spin-bond}
\end{equation}
where ${\bf f}$ is the familiar spin-Peierls order parameter
describing the distortion.  The spin part ${\bm \phi}(\{{\bf s}_i\})$
should therefore also be a doublet of the same symmetry:   
\begin{eqnarray}
\phi_1 & = & 
[({\bf s}_1 + {\bf s}_2)\!\cdot\!({\bf s}_3 + {\bf s}_4)
 - 2 {\bf s}_1 \cdot {\bf s}_2 - 2{\bf s}_3 \cdot {\bf s}_4 ]/\sqrt{12},
\nonumber
\\
\phi_2 & = & 
({\bf s}_1 - {\bf s}_2) \cdot ({\bf s}_3 - {\bf s}_4)/2.
\label{eq:f-Neel}
\end{eqnarray}
Apart from global rotations of all 4 spins, these two variables
uniquely determine the relative directions of the 4 vectors ${\bf
s}_i$ satisfying the constraint $\sum_{i=1}^4 {\bf s}_i = 0$.  It is
convenient to separate the direction and length of the spin averages
in the free energy: ${\bm \phi} = s^2 \hat{\bm \phi}$.  The free energy
then takes the following form:
\begin{equation}
F(s, {\bm \phi}) = a_0 s^2 + a_2({\bf f} \!\cdot\! \hat{\bm \phi}) s^2
+ b_0 s^4 + b_1 (\hat{\bm \phi} \!\cdot\! \hat{\bm \phi}) s^4.
\label{eq:f-Neel2}
\end{equation}
The last term is simply Eq.~(\ref{eq:biquad}); we have also dropped
the term $(\sum_{i=1}^4 {\bf s}_i)^2$ assuming that the minimization
is done over antiferromagnetic states.

Minimization can now be done separately over the length $s$ and
direction variables $\hat{\bm \phi}$.  The minimization with respect
to $s$ at a fixed $\hat{\bm \phi}$ is straightforward giving a minimum
\begin{equation}
F({\bm \phi}) = \inf_{s}{F(s, {\bm \phi})} 
= -\frac{[a_0 + a_2 ({\bf f} \!\cdot\! \hat{\bm \phi})]^2}
{2 [b_0 + b_1 (\hat{\bm \phi} \!\cdot\! \hat{\bm \phi})]}.
\end{equation}
The minimization with respect to $\hat{\bm \phi}$ is done over the
triangular domain shown in Fig.~\ref{fig:f}.  The outcome is decided
by a competition between the biquadratic exchange (\ref{eq:biquad})
and the coupling to the spin-Peierls order (\ref{eq:spin-bond}).
Generally, a distortion ${\bf f}$ pulls the vector $\hat{\bm \phi}$ in
the same direction (if $a_0<0$ and $a_2<0$), whereas the biquadratic
coupling attempts to minimize ($b_1>0$) or maximize ($b_1<0$) its
length.

\begin{figure}
\epsfig{file=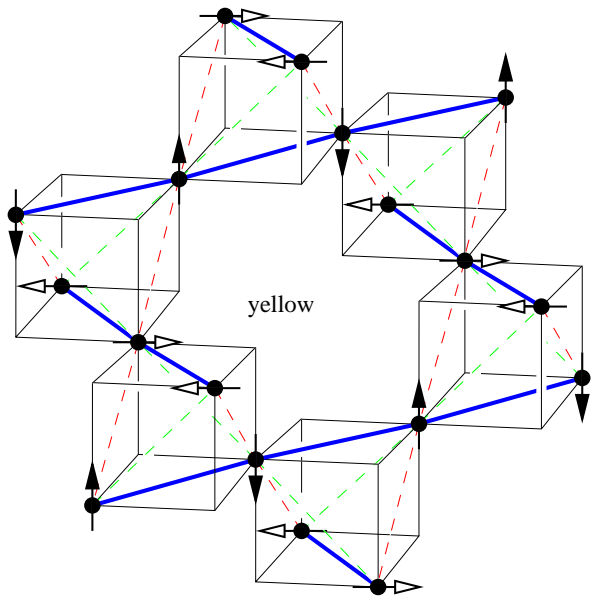,width=0.9\columnwidth}
\caption{A ${\bf q} = 0$ N\'eel state with orthogonal spins.  The
distortion is created by the same $E_g$ phonon as in
Fig.~\ref{fig:Neel-g} but with the opposite sign (blue bonds are
{\em enhanced}).}
\label{fig:Neel-g2}
\end{figure}

Fig.~\ref{fig:Neel-GL} shows the appropriate phase diagram of the
antiferromagnetic ordering for the case of a uniform tetragonal
distortion.  When the influence of the distortion dominates, we find
two N\'eel phases.  For a distortion that produces four strong bonds
per tetrahedron, the spins are collinear (e.g., ${\bf s}_1 = {\bf s}_2
= -{\bf s}_3 = -{\bf s}_4$); a distortion of the opposite sense (four
weak bonds) stabilizes a coplanar state with two {\em orthogonal}
pairs of spins (e.g., ${\bf s}_1 = -{\bf s}_2$, ${\bf s}_3 = -{\bf
s}_4$, while ${\bf s}_1 \!\cdot\! {\bf s}_3 = 0$,
Fig.~\ref{fig:Neel-g2}).  In an intermediate state, where the
biquadratic coupling dominates, the spins are no longer coplanar.  For
a positive biquadratic coupling $b_1$, they gradually interpolate
between the collinear and orthogonal orientations as the strength of
the distortion varies.  In the shaded region ($b_1<0$), a uniform
tetragragonal phase becomes unstable: the distortion acquires an
orthorombic component at the onset of the N\'eel order; in addition,
spin averages ${\bf s}_i$ break the translational symmetry of the
crystal, as in Fig.~\ref{fig:Neel-u}.


\section{Conclusion}
\label{sec:conclusion}

In this paper, we have presented a theory of the spin-Peierls
transition in a frustrated magnetic system, the Heisenberg
antiferromagnet on the ``pyrochlore'' lattice.  Several aspects
distinguish this effect from its counterpart in spin chains (e.g.,
CuGeO$_3$).  (1) The magnetic system is manifestly three-dimensional,
initially possessing a cubic symmetry.  (2) The effect is classical:
the quantum mechanics of spins plays no significant role.  (3) The
order parameter (disparity of spin correlations) has rather nontrivial
symmetry properties: its components form an irreducible doublet of the
tetrahedral symmetry group.  

What drives this spin-Peierls transition?  From the kinematical
viewpoint, the color variables (\ref{eq:f}) can be viewed as
coordinates in the manifold of ground states.  The onset of a bond
order thus lifts the large accidental degeneracy of the ground state.
The dynamical reason for the transition is the Jahn-Teller effect
occuring in the building blocks of the ``pyrochlore'' lattice,
tetrahedra of magnetic ions.  Unlike in many cubic spinels, where the
Jahn-Teller distortion is caused by an {\em orbital} degeneracy, here
it is driven by the {\em spin} degrees of freedom.  An isolated
tetrahedron (whether with quantum or classical spins) undergoes a
tetragonal distortion along one of the three orthogonal symmetry axes,
producing four strong and two weak bonds --- or vice versa, depending
on the sense of the distortion.  (The residual threefold degeneracy of
the spin-and-phonon ground state relates this problem to the 3-state
Potts model.)

A study of the spin-driven distortion of an isolated tetrahedron
provides clues to the simplest description of the spin-Peierls effect
on the lattice of corner-sharing tetrahedra, in particular the form of
the order parameter.  Depending on the parity of the phonon
responsible for the transition, Landau theory predicts a first or
second-order spin-Peierls transition.  It is gratifying to see that
these predictions are consistent with known properties of the 3-state
Potts model on the diamond lattice, onto which our problem can be
mapped.  

Although a distortion of the lattice reduces the geometric
frustration, there is no reason to expect that it induces a N\'eel
order immediately at $T_c$.  With the exception of strongly
discontinuous spin-Peierls transitions, spin order will will generally
set in at a lower temperature than bond order.  A transition between a
spin-Peierls and N\'eel phases has been analyzed above, again in the
framework of Landau theory, which predicts collinear, coplanar,
and more general antiferromagnetic orders at the lowest temperatures.

Are there experimental realizations of the bond-ordering transition
described in this work?  The spinel compound ZnCr$_2$O$_4$, in which
Cr$^{3+}$ ions ($S=3/2$) form the tetrahedral network, appears to be a
good candidate for a Heisenberg antiferromagnet on the ``pyrochlore''
lattice.  Indeed, it exhibits a magnetic transition at $T_c = 12$ K
accompanied by a structural distortion.\cite{Niziol,SHLee} N\'eel
order is observed at $T\leq T_c$, which is not surprising given that
the transition is discontinuous.  Another spinel, MgV$_2$O$_4$, shows
a sequence of two transitions:\cite{Mamiya} a structural one occurs at
$T_{c2} = 65$ K, N\'eel order appears at $T_{c1} = 42$ K.  However,
because of an orbital degeneracy (the outer electrons in V$^{3+}$ are
$3s^2p^6d^2$), the upper transition may well be triggered by the
ordinary Jahn-Teller effect common to spinels.  Therefore, we cannot
positively identify it as a spin-Peierls transition.  Nevertheless,
the N\'eel order in this compound (and also in YMn$_2$) agrees with
the prediction of a simple magnetoelastic model given here
(Fig.~\ref{fig:Neel-u}).  Lastly, a recently discovered second-order
structural phase transition\cite{Takigawa,Gaulin-cad} in the metallic
pyrochlore Cd$_2$Re$_2$O$_7$ at $T_c = 194$ K may turn out to be a
closely related Peierls transition, whose symmetry properties are
identical.

We have presented a bare-bones theoretical description of the
spin-Peierls transition in a ``pyrochlore'' antiferromagnet.  Only the
simplest ordering patterns have been discussed, namely those that do
not break the translational symmetry of the crystal.  An obvious
extension of this work would be to include bond and spin orders at
nonzero commensurate wave vectors.  E.g., a phonon with ${\bf q =
(\pi,\pi,\pi)}$ appears to be responsible for the distortion in
ZnCr$_2$O$_4$.  

\section*{Acknowledgment}

It is a pleasure to thank G. Aeppli, C. Broholm, R. J. Cava, C. L. Henley, and
S.-H. Lee for useful discussions.  The work was supported in part by
the NSF grant No. DMR-9978074 and by the David and Lucille Packard
Foundation.


\appendix

\section{A model with a coplanar ground state}
\label{sec:coplanar}

\begin{figure}
\epsfig{file=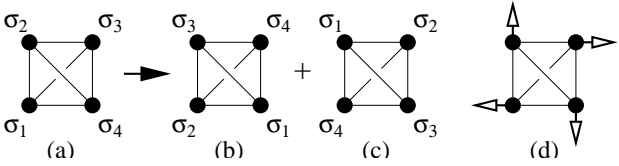,width=0.95\columnwidth}
\caption{(a--c) Cyclic exchange of 4 spins.  (d) A ground state 
with two orthogonal pairs of (classical) spins.}
\label{fig:cyclic}
\end{figure}

In addition to pairwise spin exchange, which gives rise to the
Heisenberg interaction ${\bf S}_j \cdot {\bf S}_j$, spins can be
involved in longer exchange cycles, such as $123 \mapsto 231$ and 312.
The cyclic exchange of 3 spins induces the same pairwise Heisenberg
interaction, which has already been considered.  The next nontrivial
contribution comes from 4-spin cyclic exchange
\begin{eqnarray*}
P_{1234} &=& ({\bf S}_1 \cdot {\bf S}_3) ({\bf S}_2 \cdot {\bf S}_4) 
+ ({\bf S}_1 \cdot {\bf S}_4) ({\bf S}_2 \cdot {\bf S}_3) 
\\
&-& ({\bf S}_1 \cdot {\bf S}_2) ({\bf S}_3 \cdot {\bf S}_4),
\end{eqnarray*}
which --- for $S=1/2$ --- moves spin states clockwise or
counterclockwise around the loop 1234 (Fig.~\ref{fig:cyclic}).  For
localized spins, this interaction is weaker than pairwise exchange and
can be considered as a perturbation.  Nevertheless, its signature has
apparently been detected\cite{Aeppli} in the spin-wave spectrum of
La$_2$CuO$_4$.

A tetrahedron has three loops of length four: 1234, 1324, and 1243.  
Quantities $P_{1234}$, $P_{1324}$, and $P_{1243}$ contain the 
trivial representation $A_1$ and the doublet $E$.  

The sum $P_{1234} + P_{1324} + P_{1243}$, being invariant under all
symmetry operations, can enter an expression for the energy on its
own.  In the subspace of the ground states of the Heisenberg
Hamiltonian, this perturbation can be written as a biquadratic term 
\[
J_4\sum_{i>j} ({\bf S}_i \cdot {\bf S}_j)^2 = \frac{3J_4 f^2}{2} +
\mbox{ const}.
\]
The outcome depends on the sign of the coupling constant $J_4$.  If
the constant is negative, it prefers the largest magnitude of ${\bf
f}$ and selects the three collinear spin states (primary colors in
Fig.~\ref{fig:f}).  A positive $J_4$ suppresses ${\bf f}$, selecting
the white state with all bonds equivalent; spin averages make angles
of $\arccos{(-1/3)} \approx 109^\circ$ with each other.

The two differences between $P_{1234}$, $P_{1324}$, and $P_{1243}$
form an irreducible doublet $E$.  They can therefore couple to the
phonon doublet of the same symmetry adding this magnetoelastic term to
the energy:
\[
J_4' \left( 
\frac{P_{1324} + P_{1243} - 2P_{1234}}{\sqrt{6}} x_1
+ \frac{P_{1324} - P_{1243}}{\sqrt{2}} x_2
\right),
\]
where $J_4'$ describes variation of the cyclic exchanges under a
tetragonal or orthorombic distortion.  After adding an elastic term $k
x^2/2$ and minimizing with respect to the phonon variables ${\bf x}$,
one obtains the following contribution to the energy:
\begin{eqnarray}
- \ \frac{2{J_4'}^2}{3k} \big\{ \!\!\!\!\!\!\!\!\!
&&\left[ ({\bf S}_1 \!\cdot\! {\bf S}_2)^2 
- ({\bf S}_2 \!\cdot\! {\bf S}_3)^2 \right]^2
\nonumber
\\
&+& \left[ ({\bf S}_2 \!\cdot\! {\bf S}_3)^2 
- ({\bf S}_3 \!\cdot\! {\bf S}_1)^2 \right]^2
\\
&+& \left[ ({\bf S}_3 \!\cdot\! {\bf S}_1)^2 
- ({\bf S}_1 \!\cdot\! {\bf S}_2)^2 \right]^2
\big\}.
\nonumber
\end{eqnarray}
The energy is lowered by the greatest amount in coplanar states with
spins making angles of $90^\circ$ and $180^\circ$ with one another.
These states have two pairs of frustrated bonds and are marked in
secondary colors in Fig.~\ref{fig:f}: e.g. a state with blue and red 
frustrated bonds is magenta.

As the two physical forces described above --- the four-spin exchange
and its coupling to the phonons --- favor different ground states, the
outcome is decided by their relative strengths.  In particular, the
ground state has orthogonal spins [Fig.~\ref{fig:cyclic}(d)] when the
4-spin exchange is very sensitive to atomic displacements (and
therefore the distortion is large).  The reason for this effect can be
understood as follows.  A strong tetragonal distortion enhances the
4-spin exchange around one loop and suppresses the same around the
remaining two.  When the latter are completely switched off, we are
dealing with a square, and for a square the 4-spin cyclic exchange
produces a ground state with orthogonal spins.\cite{Chubukov}

\section{Single tetrahedron: quantum spins}
\label{sec:quantum}

Quantum spins coupled to classical distortions of the tetrahedron have
the following Hamiltonian in the subspace of ground states:
\begin{equation}
H = - J' \, {\bf f \! \cdot \! x} + k\, x^2/2,
\label{eq:h-x-f}
\end{equation}
which is formally the same as the classical energy (\ref{eq:e-x-f}).
For a fixed ``direction'' of the distortion 
\[
\hat{\bf n} = {\bf x}/x = (\cos{\alpha}, \sin{\alpha}), 
\] 
the operator ${\bf f} \cdot \hat{\bf n} = f_1 \cos{\alpha} + f_2
\sin{\alpha}$ has $2S+1$ eigenvalues $\lambda_\sigma$, $\sigma=0
\ldots 2S$ in the ground-state manifold.  The Hamiltonian
(\ref{eq:h-x-f}) has the following energy levels:
\[
E_\sigma(x, \hat{\bf n}) = -J' \lambda_\sigma x + k x^2/2.
\]
Minimization with respect to the magnitude of the distortion $x$ gives
the following result:
\[
E_{\sigma}(\hat{\bf n}) = \inf_x E_\sigma(x, \hat{\bf n}) 
= -{J'}^2 \lambda_\sigma^2/2k,
\]
The energy is lowest in a spin state with the largest (in absolute
terms) eigenvalue $\lambda_\sigma$ of the operator ${\bf f} \cdot
\hat{\bf n}$.  The final step is minimization with respect to the
direction $\hat{\bf n}$.

\begin{figure}
\epsfig{file=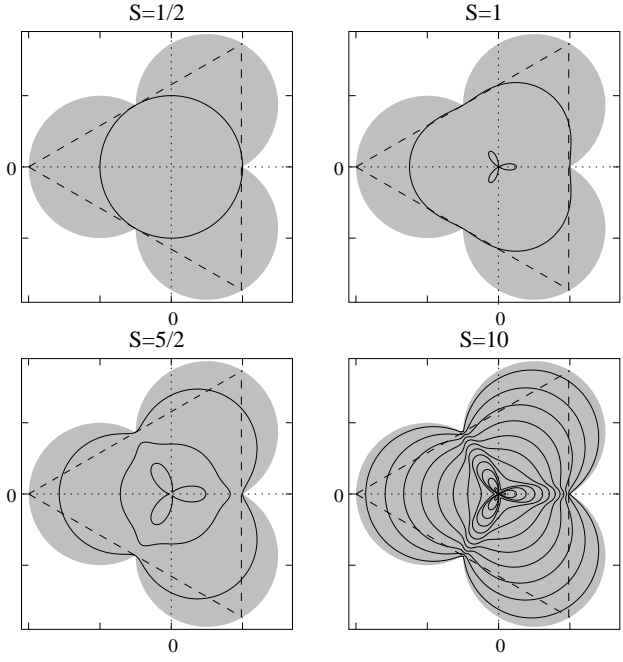,width=\columnwidth}
\caption{Solid lines: eigenvalues $\lambda_\sigma$ of the operator
${\bf f} \!\cdot\!  \hat{\bf n} = f_1 \cos{\alpha} + f_2 \sin{\alpha}$
in polar coordinates $(\lambda_\sigma,\alpha)$ for several spin
lengths $S$.  Shaded area: the classical result for $S\to\infty$ 
(properly rescaled).}
\label{fig:fnn}
\end{figure}

Let us illustrate the minimization procedure using the classical
problem of Section \ref{sec:jt-tetrahedron} as an example.  To this
end, we show classically allowed values of the vector $\lambda
\hat{\bf n} = ({\bf f} \! \cdot \! \hat{\bf n})\, \hat{\bf n}$ as
shaded areas in Fig.~\ref{fig:fnn} (the values of ${\bf f}$ fill the
interior of the regular triangle shown in dashed lines).  The largest
magnitude $|\lambda \hat{\bf n}| = |\lambda|$ is found in the
directions $\alpha = 0, \pm 2\pi/3$, which correspond to the three
collinear states.  In the quantum case, $\lambda_\sigma$ has a
discrete spectrum, therefore allowed values of $\lambda_\sigma
\hat{\bf n}$ will show as lines on the same graph.

The main subtlety of the quantum problem is the noncommutativity of
the bond operators $f_1$ and $f_2$:
\begin{equation}
[f_1, f_2] = -2\sqrt{3}\, i\, \chi, 
\end{equation}
where $\chi$ is the operator of chirality
\[
\chi = {\bf S}_1 \cdot ({\bf S}_2 \times {\bf S}_3) = 
- {\bf S}_1 \cdot ({\bf S}_2 \times {\bf S}_4) = \ldots
\]
Therefore eigenvalues of ${\bf f} \cdot \hat{\bf n}$ cannot be
constructed from those of $f_1$ and $f_2$, but rather must be determined
for every direction $\hat{\bf n}$.

\subsection{Matrix elements of the operator ${\bf f}$}
\label{sec:quantum-diag}

In the $(2S+1)$-dimensional subspace of singlet ground states we choose
the basis $\{|\sigma\rangle\}$, $\sigma = 0 \ldots 2S$ being the total
spin of pair 12.  (The spin of pair 34 must be the same in order to
form a total spin of 0.)  The operator $f_1$ is also diagonal in this
basis because
\[
f_1 = [{\bf S}_{12} \cdot {\bf S}_{34} - {\bf S}_{12}^2 
- {\bf S}_{34}^2 + 4S(S+1)]/\sqrt{12}.
\]
Its eigenvalues are
\begin{equation}
f_1 = [4S(S+1)-3\sigma(\sigma+1)]/\sqrt{12}. 
\end{equation}

The operator $f_2$ is off-diagonal.  To compute its matrix elements, 
write out an expression for the singlet $|\sigma\rangle$:
\[
|\sigma\rangle = \frac{1}{\sqrt{2\sigma+1}} \sum_{\mu=-\sigma}^{\sigma}
(-1)^{\sigma-\mu} \ |\sigma,\mu\rangle_{12} \ |\sigma,-\mu\rangle_{34}. 
\]
Here $|\sigma,\mu\rangle_{12}$ is the state of pair 12 with total spin
$\sigma$ and its projection $\mu$ onto a chosen axis.  Then, by
definition,
\begin{widetext}
\begin{eqnarray}
\langle\sigma'| f_2 |\sigma\rangle 
&=& \frac{1}{2\sqrt{(2\sigma'+1)(2\sigma+1)}} 
\sum_{\mu'=-\sigma'}^{\sigma'}
\sum_{\mu=-\sigma}^{\sigma}
(-1)^{\sigma+\sigma'-\mu-\mu'} \ 
\langle\sigma',-\mu'| ({\bf S}_3-{\bf S}_4) |\sigma,-\mu\rangle_{34} 
\cdot
\langle\sigma',\mu'| ({\bf S}_1-{\bf S}_2) |\sigma,\mu\rangle_{12}
\nonumber
\\
&=& \frac{1}{2\sqrt{(2\sigma'+1)(2\sigma+1)}} 
\sum_{\mu'=-\sigma'}^{\sigma'}
\sum_{\mu=-\sigma}^{\sigma}
\langle\sigma,\mu| ({\bf S}_1-{\bf S}_2) |\sigma',\mu'\rangle_{12}
\cdot
\langle\sigma',\mu'| ({\bf S}_1-{\bf S}_2) |\sigma,\mu\rangle_{12}.
\label{eq:matrix-element}
\end{eqnarray}
\end{widetext}
In simplifying this expression, we have replaced indices 3 and 4 with
2 and 1 because the matrix elements involve one pair at a time.  Also,
we have used the properties of time reversal\cite{LL} to simplify the
first matrix element in the summand.  

The last line of Eq.~(\ref{eq:matrix-element}) contains matrix elements
of ${\bf S}_1-{\bf S}_2$, which is (a) a vector; (b) antisymmetric in
$1\leftrightarrow 2$.  These lead to a selection rule: 
\begin{equation}
\langle\sigma'| f_2 |\sigma\rangle = 0 
\mbox{ unless }
\sigma' = \sigma \pm 1.
\label{eq:selection-rule}
\end{equation}
Completeness of the basis $\{|\sigma',\mu'\rangle\}$ allows us to 
further simplify the right-hand side at the expense of producing a set
of $2S+1$ coupled equations.  This is done by summing over $\sigma'$
with appropriate weights:
\begin{widetext}
\begin{eqnarray*}
\sum_{\sigma'=0}^{2S} \sqrt{\frac{2\sigma'+1}{2\sigma+1}} 
\langle\sigma'| f_2 |\sigma\rangle 
=  \frac{1}{4\sigma+2} 
\sum_{\mu=-\sigma}^{\sigma}
\langle\sigma,\mu| ({\bf S}_1-{\bf S}_2)^2 |\sigma,\mu\rangle_{12}
= 2S(S+1) - \frac{1}{2}\sigma(\sigma+1).
\label{eq:sum-rule}
\end{eqnarray*}
\end{widetext}
Solving these gives the matrix elements of $f_2$. 

The operator ${\bf f}$ has the following nonzero matrix elements:
\begin{eqnarray}
\langle \sigma| f_1 |\sigma \rangle = 
\frac{4S(S+1)-3\sigma(\sigma+1)}{\sqrt{12}}, \\
\langle \sigma-1| f_2 |\sigma \rangle = 
\langle \sigma| f_2 |\sigma-1 \rangle = 
\frac{\sigma[(2S+1)^2-\sigma^2]}{2\sqrt{4\sigma^2-1}}.
\end{eqnarray}

\subsection{Eigenvalues of the operator ${\bf f} \!\cdot\! \hat{\bf n}$}

For $S=1/2$, we have
\begin{equation}
{\bf f} \!\cdot\! \hat{\bf n} = 
\frac{\sqrt{3}}{2}
\left( \begin{array}{rr}
\cos{\alpha} & \sin{\alpha} \\
\sin{\alpha} & -\cos{\alpha}
\end{array} \right).
\end{equation}
The eigenvalues $\lambda = \pm \sqrt{3}/2$ are independent of the
direction $\hat{\bf n}$.  Thus there is no preferred direction for the
distortion, unless one introduces some additional, nonlinear
couplings.  This fact was noted previously by Yamashita and
Ueda.\cite{Ueda}

For $S=1$, 
\begin{equation}
{\bf f} \!\cdot\! \hat{\bf n} = 
\frac{1}{\sqrt{3}}
\left( \begin{array}{ccc}
4\cos{\alpha} & 4\sin{\alpha} & 0\\
4\sin{\alpha} & \cos{\alpha} & \sqrt{5}\sin{\alpha} \\
0 & \sqrt{5}\sin{\alpha} & -5 \cos{\alpha}
\end{array} \right).
\end{equation}
Its eigenvalues are given by the equation
\[
\lambda^3 - 7\lambda + \frac{20\cos{3\alpha}}{3\sqrt{3}} = 0.
\]
The largest eigenvalue is attained when $\alpha = \pi$ or $\pm\pi/3$,
i.e. when $\hat{\bf n}$ points towards one of the corners of the
triangle (the three tetragonal distortions).  For $\theta = \pi$, the
total spin of bond 12 is $\sigma=2$ (the same goes for bond 34).  For the
other choices of $\alpha$, the highest spin is found on other pairs of
opposite bonds.

We have checked higher spin values $S \leq 10$ and always found that
the lowest energy is obtained for a tetragonal distortion, when two
opposite bonds have the highest spin $2S$ and are thus most strongly
frustrated.  One can use perturbation theory to show that $\alpha =
\pi$ is at least a local maximum of $|\lambda_{2S}|$ for any $S>1/2$:
\[
\frac{\lambda_{2S}(\alpha)}{\lambda_{2S}(0)} 
= 1 - \frac{2S-1}{4S-1}\alpha^2 + {\cal O}(\alpha^4).
\]
As the value of $S$ increases, eigenvalues $\lambda_\sigma$ fill the
classical region (three overlapping shaded circles in
Fig.~\ref{fig:fnn}).



\begin{thebibliography}{99}
\bibitem{Wannier} G. H. Wannier, Phys. Rev. {\bf 79}, 357 (1950). 
\bibitem{Houttapel} R. M. F. Houttapel, Physica {\bf 16}, 425 (1950).
\bibitem{SR} P. Schiffer and A. P. Ramirez, Comments
Cond. Mat. Phys. {\bf 18}, 21 (1996).
\bibitem{Moessner} R. Moessner, cond-mat/0010301.  
\bibitem{MC} R. Moessner and J. T. Chalker, \prl {\bf 80}, 2929 (1998);
\prb {\bf 58}, 12049 (1998).
\bibitem{henley-unpub} C. L. Henley (unpublished).
\bibitem{HBB} A. B. Harris, A. J. Berlinsky, and C. Bruder,
J. Appl. Phys. {\bf 69}, 5200 (1991).
\bibitem{CL} B. Canals and C. Lacroix, \prl {\bf 80}, 2933 (1998).
\bibitem{Ueda} Y. Yamashita and K. Ueda, \prl {\bf 85}, 4960 (2000).
\bibitem{TMS} O. Tchernyshyov, R. Moessner, and S. L. Sondhi, 
\prl {\bf 88}, 067203 (2002).
\bibitem{endnote1} Of course it is possible that both mechanisms are
preempted by ordering coming from other, purely magnetic terms, beyond
the nearest-neighbor model.
\bibitem{Kittel} Biquadratic exhange was first described in the
context of magnetoelastic interactions by C. Kittel, Phys. Rev. {\bf 120}, 
335 (1960).
\bibitem{Potts} F. Y. Wu, \rmp {\bf 54}, 235 (1982).
\bibitem{Ono} I. Ono, Prog. Theor. Phys. Suppl. {\bf 87}, 102 (1986).
\bibitem{Banavar} J. R. Banavar, G. S. Grest, and D. Jasnow, \prl {\bf 45}, 
1424 (1980).
\bibitem{Rosengren} A. Rosengren and S. Lapinskas, \prl {\bf 71}, 165 (1993).
\bibitem{Rahman} S. Rahman, E. Rush, and R. H. Swendsen, \prb {\bf 58}, 
9125 (1998). 
\bibitem{Lapinskas} S. Lapinskas and A. Rosengren, \prl {\bf 81}, 1302 (1998).
\bibitem{Niziol} S.~Nizio\l, Phys. Status Solidi A {\bf 18}, K11 (1973).
\bibitem{SHLee} S.-H. Lee, C. Broholm, T.H. Kim, W. Ratcliff II, and 
S.-W. Cheong, \prl {\bf 84}, 3718 (2000).
\bibitem{Mamiya} H. Mamiya, M. Onoda, T. Furubayashi, J. Tang, and
I. Nakatani, J. Appl. Phys. {\bf 81}, 5289 (1997).
\bibitem{Takigawa} O. Vyaselev, K. Arai, K. Kobayashi, J. Yamazaki,
K. Kodama, M. Takigaawa, M. Hanawa, and Z. Hiroi, {\tt
cond-mat/0201215}.
\bibitem{Gaulin-cad} J.P. Castellan, B.D. Gaulin, J. van Duijn,
M.J. Lewis, M.D. Lumsden, R. Jin, J. He, S.E. Nagler, and D. Mandrus, 
{\tt cond-mat/0201513}.  
\bibitem{Aeppli} R. Coldea, S. M. Hayden, G. Aeppli, T. G. Perring,
C. D. Frost, T. E. Mason, S.-W. Cheong, and Z. Fisk, 
\prl {\bf 86}, 5377 (2001).  
\bibitem{Chubukov} A. V. Chubukov, E. Gagliano, and C. Balseiro, \prb
{\bf 45}, 7889-7898 (1992).
\bibitem{LL} L. D. Landau and E. M. Lifshitz, {\em Quantum Mechanics}.

\end{thebibliography}
\end{document}